\documentclass[preprint,prc,aps,nofootinbib,superscriptaddress]{revtex4}
\usepackage{graphicx}
\usepackage{epsfig}
\usepackage{bm}
\usepackage{amssymb}

\newcommand{\lag}{\mathcal L}

\begin{document}

\title{Neutron stars with the Bose-Einstein condensation of antikaons as MIT Bags}

\author{C. Y. Ryu}
%\footnote{Present address : Department of Physics,
%Soongsil University, Seoul 156-743}
\email{cyryu@skku.edu}
\affiliation{Department of Physics and Institute of Basic Science,
Sungkyunkwan University, Suwon 440-746}

\author{C. H. Hyun}
%\email{hch@daegu.ac.kr}
\affiliation{Department of Physics Education, Daegu University,
Gyeongsan 712-714}

\author{S. W. Hong}
%\email{swhong@skku.ac.kr}
\affiliation{Department of Physics and Institute of Basic Science,
Sungkyunkwan University, Suwon 440-746}

\date{December 26, 2008}

\begin{abstract}
We investigate the properties of an antikaon in medium,
regarding it as a MIT bag. We first construct the MIT bag model for a kaon
with $\sigma^*$ and $\phi$ in order to describe
the interaction of s-quarks in hyperonic matter in the framework of the
modified quark-meson coupling model. The coupling constant $g'^{B_K}_\sigma$
in density dependent bag constant $B(\sigma)$ is treated as a free parameter
to reproduce the optical potential of a kaon in symmetric matter, and
all other couplings are determined by SU(6) symmetry and the
quark counting rule. With various values of the kaon potential,
we calculate the effective mass of a kaon in medium to compare it
with that of a point-like kaon.
We then calculate the population of octet baryons, leptons and $K^-$,
and the equation of state for the neutron star matter.
The results show that kaon condensation in hyperonic matter
is sensitive to the s-quark interaction and
how to treat the kaon.
The mass and radius of a neutron star are obtained by
solving Tolmann-Oppenheimer-Volkoff equation.
\end{abstract}

\maketitle

\section{Introduction}
%234567890123456789012345678901234567890123456789012345678901234567890
The properties of an antikaon in medium are attracting much interests recently.
Since a theoretical calculation \cite{akaishi} predicted
deeply bound $K^-$ states in $K^-pp$, $K^-ppn$ and $K^-pnn$ systems,
many studies have been reported in both theory and experiment.
The KEK group \cite{suzuki} reported the observation of $S^0(3115)$
and $S^+(3140)$ states, interpreting them as deeply bound states of $K^-$,
though in a more recent experiment \cite{sato08} with a better statistics
the existence of $S^0$ state could not be confirmed.
FINUDA group also reported the observation of a deeply bound state
for $K^-pp$ with binding energy $B=115 \pm 9$ MeV
and width $\Gamma = 67 \pm 16$ MeV \cite{finuda}.
On the other hand, the authors in Refs. \cite{oset,magas} claimed
that the observed state could be explained in terms of
the final state interaction of produced $\Lambda p$ pairs.
Therefore, further studies
are needed both theoretically and experimentally to confirm such states.
If a deeply bound kaonic state exists indeed, it can be formed
by very strong attraction between nucleons and an antikaon,
which may lead to kaon condensation in dense nuclear matter.

Possible existence of kaon condensation in dense matter
was proposed about two decades ago \cite{kaplan}.
The subject has been studied with various models such as chiral perturbation theory
\cite{kaplan,brown,brown2,muto}, relativistic mean field models like
quantum-hadrodynamics (QHD) \cite{glendening,glen-prc,banik}
and quark-meson coupling (QMC) model \cite{menezes,cyryu07}.
In relativistic mean field approach, the interaction among baryons is
described by scalar ($\sigma$) and vector ($\omega$) meson fields
mediating attractive and repulsive forces, respectively.
%The interaction between nucleons and antikaons can also be
%described in such an approach \cite{glendening},
In the mean field approach, due to the antiparticle nature,
an antikaon feels strong attraction by both scalar and vector meson
fields, which determines the optical potential of an antikaon.
The kaon condensation in medium is known to be sensitive to the value of
the optical potential.
When we consider an antikaon in medium,
it can be treated as a point-like particle
in view of its property as a pseudo-Goldstone boson.
However, applying the OZI rule, which says the s-quark does not interact
with $\sigma$ and $\omega$ mesons, the optical potential of an antikaon
depends only on the interaction of $\bar u$,
which can feel strong attraction due to the exchange of $\sigma$ and
$\omega$ mesons.
Therefore it is worthwhile to treat an antikaon as a MIT bag and
compare the values of physical observables with
those obtained from a point-like kaon.

The QMC model originally developed by Guichon \cite{guichon} assumes
that quarks inside baryon bags interact with
each other through the exchange of $\sigma$, $\omega$ and $\rho$ meson fields.
The model has been further developed by several authors
\cite{fleck,jin,saito-npa,saito98,cyryu05}.
An important revision of the model was made in Ref. \cite{jin}
by introducing a density dependent bag constant to simulate
partial deconfinement of quarks at high densities and to get
the meson field strengths as predicted by relativistic phenomenology.
This model was called the modified QMC (MQMC) model.
Then a model for a kaon as a MIT bag was proposed in Ref. \cite{saito98}
in the framework of QMC model in asymmetric matter,
in which the interaction between nucleons and a kaon
was mediated by $\sigma$, $\omega$ and $\rho$ meson fields.
The kaon condensation in hyperonic nuclear matter was then studied in Ref. \cite{menezes}
with the QMC model, including $\sigma$, $\omega$ and $\rho$ meson fields,
but the interaction
between s-quarks and the density dependence of the bag constant were
not considered.

In this work, we extend the previous models \cite{saito98,menezes}
to further include the interaction of s-quarks through the exchange of
$\sigma^*$ and $\phi$ meson fields,
and assume the density dependence of the bag constants.
We then study the Bose-Einstein condensation of $K^-$ in the neutron star
with the density dependent bag constant for a kaon bag as
in Ref. \cite{cyryu05} in the framework of the MQMC model.
The coupling constant $g'^{B_K}_\sigma$ for the density dependent
bag constant of a kaon is regarded as a free parameter
to be determined from the kaon optical potential,
which we choose as $U_K = -120$, $-140$, $-160$ MeV.
Our results show that the EoS with kaon condensation
differs significantly depending on whether we treat a kaon
as a point-like particle or a MIT bag.
Because of the repulsion due to the s-quark interaction at high
densities, the equation of states (EoS) from the QHD and QMC models
without s-quark interactions are softer than those from the models
including the s-quark interaction.
We find that the interaction between s-quarks,
which was not included in Ref. \cite{menezes} can make the EoS
quite different.

In Section II, we present our model, extending
the previous QMC model \cite{saito98} for a kaon to include
the density dependence of the bag constant (i.e., MQMC model) and
the interaction between the s-quarks in hyperonic nuclear matter.
By treating a kaon as a MIT bag we calculate
the equations of motion for mesons with kaon condensation.
Energy density and pressure of the neutron star matter with kaon bags
are obtained.

In Section III, we show the numerical results for the EoS and population profile
of neutron star matter with hyperons at densities up to ten times normal density.
Using the calculated EoS, we obtain
the mass-radius relation of neutron star with hyperons and kaons.
Summary follows in Section IV.

\section{Models}
\subsection{Kaon and antikaon as MIT bags in hyperonic matter}
In the (M)QMC model, baryons are regarded as MIT bags and the interaction
between the baryons is mediated by the exchange of meson fields among quarks
inside the baryon bags. As mentioned in the Introduction, we treat a kaon
as a MIT bag in this work.
Thus we use the same framework for both kaons and baryons,
treating all virtual meson fields as point-like particles.
In Ref. \cite{saito98} a kaon and an antikaon as MIT bags in nuclear matter were
assumed to interact with nucleons through the interaction between the $u$- and $d$-quarks
and $\sigma$, $\omega$ and $\rho$ meson fields, while s-quarks were assumed
to be non-interacting with any particles.
Here, we extend the model of Ref. \cite{saito98} to include the interaction between
the s-quarks inside the kaon and baryon bags
through $\sigma^*$ and $\phi$ meson fields.
To show explicitly how kaon bags interact with five meson fields,
we start with the Dirac equations for quarks and antiquarks in mean field approximation
\begin{eqnarray}
[ i \gamma \cdot \partial - (m_u - g^q_\sigma \sigma) \mp
         \gamma^0 (g^q_\omega \omega_0 + \frac 12 g^q_\rho \rho_{03}) ]
         { u \choose \bar u } = 0
\end{eqnarray}
\begin{eqnarray}
[ i \gamma \cdot \partial - (m_d - g^q_\sigma \sigma) \mp
         \gamma^0 (g^q_\omega \omega_0 - \frac 12 g^q_\rho \rho_{03}) ]
         { d \choose \bar d } = 0
\end{eqnarray}
\begin{eqnarray}
[ i \gamma \cdot \partial - (m_s - g^q_{\sigma^*} \sigma^*) \mp
         \gamma^0 g^q_\phi \phi_0 ]
         { s \choose \bar s} = 0,
\end{eqnarray}
where $\omega_0$ and $\phi_0$ are, respectively,
the time components of $\omega$ and $\phi$ meson fields and
$\rho_{03}$ is the z-component of the time-component of $\rho$ meson field.
For u and d-quarks we use $m_u = m_{\bar u} = m_d = m_{\bar d} = 0$,
while for s and $\bar s$ quarks, $m_s = m_{\bar s} = 150$ MeV is used.
In the assumption that all the quarks are in the ground state,
we can use the following normalized, static solution for the quarks in a kaon bag,
\begin{eqnarray}
\psi_i ( \vec r, t ) = N_i e^{i \epsilon_i t / R_K} \phi_i (\vec r),
~~~~i = u,~ \bar u, ~d, ~\bar d, ~s, ~\bar s
\end{eqnarray}
where $N_i$ , $\phi_i (\vec r)$ and $R_K$ are the normalization factor,
the spatial part of the wave function and the bag radius of a kaon bag, respectively.
The eigenenergies of quarks in units of $1/R_K$ can be obtained as
\begin{eqnarray}
{\epsilon_u \choose \epsilon_{\bar u} } &=& \Omega_u \pm
           R_K (g^q_\omega \omega_0 + \frac 12 g^q_\rho \rho_{03}) \\
{\epsilon_d \choose \epsilon_{\bar d} } &=& \Omega_d \pm
           R_K (g^q_\omega \omega_0 - \frac 12 g^q_\rho \rho_{03}) \\
{\epsilon_s \choose \epsilon_{\bar s} } &=& \Omega_s \pm
           R_K g^q_\phi \phi_0
\end{eqnarray}
where $\Omega_{q} = \sqrt{x_q^2+(R_K m_q^*)^2}$ with
$m_q^* = m_q - g^q_\sigma \sigma$ for $q=u, ~\bar u, ~d, ~\bar d$
and for $q=s$ and $\bar s$ $\Omega_s = \sqrt{x_s^2+(R_K m_s^*)^2}$
with $m_s^* = m_s - g^q_{\sigma^*}\sigma^*$.
The energies of a kaon and an antikaon can be obtained through the sum of
the energies of their quark and antiquark components as
\begin{eqnarray}
{ \omega_{K^+} \choose \omega_{K^-} } =
           m_K^* \pm (g^q_\omega \omega_0 - g^q_\phi \phi_0
            + \frac 12 g^q_\rho \rho_{03} )
 \label{kenergy1}
\end{eqnarray}
\begin{eqnarray}
{\omega_{K^0} \choose \omega_{{\bar K}^0}} =
           m_K^* \pm (g^q_\omega \omega_0 - g^q_\phi \phi_0
            - \frac 12 g^q_\rho \rho_{03})
 \label{kenergy2}
\end{eqnarray}
where the effective mass of a kaon $m_K^*$ is calculated by eliminating
the spurious motion of quarks in a bag from the bag energy \cite{fleck,cyryu05},
\begin{eqnarray}
m_K^* = \sqrt{E_K^2 - \sum_q \Big ( \frac{x_q}{R_K} \Big )^2 }.
\label{k-efmas}
\end{eqnarray}
Here the bag energy is given as
\begin{eqnarray}
E_K = \frac{\Omega_u + \Omega_s - Z_K}{R_K} + \frac 43 \pi R_K^3 B_K
\label{kmass}
\end{eqnarray}
with the bag constant $B_K$ and a phenomenological constant $Z_K$,
which are determined by the minimum condition
\begin{eqnarray}
\frac{\partial m_K}{\partial R} {\Big |}_{R=R_K} = 0
\end{eqnarray}
to produce the free mass of a kaon in vacuum with $R_K = 0.4$ fm.
In eq. (\ref{kenergy1}), one can see both $\sigma$ and $\omega$ mesons give the
attraction to an antikaon, while the $\phi$ meson fields give the repulsion.
In addition, because the value of $\rho_{03}$ is usually negative in asymmetric matter,
both $\phi$ and $\rho$ meson fields play a role to prohibit the
kaon condensation in the neutron star matter.

In the MQMC model, the baryon bag constant $B$ in medium
depends on $\sigma$ and $\sigma^*$.
For the kaon bag also we use the density dependent bag constant
given by
\begin{eqnarray}
B_K(\sigma, \sigma^*)
= B_0 \exp [- 4 {g'}_\sigma^{B_K} (\sigma + \sqrt 2 \sigma^*) / m_K ],
\end{eqnarray}
where $B_0$ and $m_K$ are the bag constant and
the mass of a kaon in vacuum, respectively.
Here the factor $\sqrt 2$ is from the SU(6) symmetry.
By applying a boost,
we can obtain a general dispersion relation
for a kaon and an antikaon with momentum $\vec k$.
We can write a Lagrangian
for a kaon and an antikaon bag interacting with meson fields as
\begin{eqnarray}
\lag_K = (D^\mu K)^\dag D_\mu K - {m_K^*}^2 K^\dag K
\label{lagk}
\end{eqnarray}
where the covariant derivative $D_\mu = \partial_\mu + i g_\omega^K \omega_\mu
-i g_\phi^K\phi_\mu + i g_\rho^K \frac{\tau_3}{2}\rho_\mu$.
For the s-wave ($\vec k = 0$), we can easily check that this Lagrangian provides
the dispersion relations in Eqs. (\ref{kenergy1}) and (\ref{kenergy2}).
Different signs in front of $\omega$ and
$\phi$ meson fields originate from the quark and
antiquark components of a kaon.

\subsection{Kaon condensation in neutron star}
The total Lagrangian for hyperonic matter with baryon octet, five meson fields, leptons and
(anti) kaons in mean field approximation is given by
\begin{eqnarray}
\lag  = \lag_{matter} + \lag_K
\end{eqnarray}
where
\begin{eqnarray}
\lag_{matter} &=& \sum_B \bar \psi_B [i\gamma \cdot \partial
   - m_B^*(\sigma, \sigma^*) +
   \gamma^0(\omega_0 + \phi_0 + \frac 12 \tau_3 \rho_{03}) ] \psi_B \\
   &+& \frac 12 m_\omega^2 \omega_0^2 + \frac 12 m_\phi^2 \phi_0^2
   + \frac 12 m_\rho^2 \rho_{03}^2
   + \sum_l \bar \psi_l ( i \gamma \cdot \partial -m_l) \psi_l,
\label{eq:lagb}
\end{eqnarray}
and $\lag_K$ is given by Eq. (\ref{lagk}).
The effective mass of a baryon as a MIT bag $m_B^*(\sigma, \sigma^*)$
can be written as \cite{fleck,greiner}
\begin{eqnarray}
m^*_B = \sqrt{E^2_B - \sum_q  \left(\frac{x_q}{R} \right)^2}
\label{eq:efmass}
\end{eqnarray}
where the bag energy of a baryon is given by
\begin{eqnarray}
E_B &=& \sum_q  \frac{\Omega_q}{R} - \frac{Z_B}{R}
+ \frac{4}{3} \pi\, R^3\, B_B.
\label{eq:bagery}
\end{eqnarray}
The bag constant $B_B$ and a phenomenological constant $Z_B$
are fitted to reproduce the free mass of each baryon
at a given bag radius $R$, respectively.
In MQMC model $B_B$ depends on the matter density
and can be written as
\begin{eqnarray}
B_B (\sigma,\, \sigma^*)  =
B_{B0} \exp \left\{ -4{g'}_\sigma^B \left(\sum_{q=u,d} n_q \sigma
+ (3-\sum_{q=u,d}n_q) \sqrt 2 \sigma^*\right) / m_B \right\},
\label{eq:bag}
\end{eqnarray}
where $m_B$ is the bare mass of the baryon $B$ and
the factor $\sqrt 2$ is from the SU(6) symmetry.

When considering the s-wave $K^-$ condensation,
the equations for five meson fields are obtained as
\begin{eqnarray}
m_\sigma^2 \sigma = \sum_B g_{\sigma B} C_B(\sigma)
\frac{2J_B+1}{2\pi^2} \int_0^{k_B} \frac{m_B^*}{[k^2+{m_B^*}^2]^{1/2}}k^2 dk
+ g_{\sigma K}C_K(\sigma) \rho_K,
\label{sigma}
\end{eqnarray}
\begin{eqnarray}
m_{\sigma^*}^2 \sigma^* = \sum_B g_{\sigma^* B} C_B(\sigma^*)
\frac{2J_B+1}{2\pi^2} \int_0^{k_B} \frac{m_B^*}{[k^2+{m_B^*}^2]^{1/2}}k^2 dk
+ g_{\sigma^* K}C_K(\sigma^*) \rho_K,
\label{sigmastar}
\end{eqnarray}
\begin{eqnarray}
m_\omega^2 \omega_0 = \sum_B g_{\omega B} (2J_B + 1) k_B^3 / (6\pi^2)
- g_{\omega K}\rho_K,
\end{eqnarray}
\begin{eqnarray}
m_\phi^2 \phi_0 = \sum_B g_{\phi B} (2J_B + 1) k_B^3 / (6\pi^2)
+ g_{\phi K}\rho_K,
\end{eqnarray}
\begin{eqnarray}
m_\rho^2 \rho_{03} = \sum_B g_{\rho B} I_{3B} (2J_B + 1) k_B^3 / (6\pi^2)
-g_{\rho K} I_{3K}\rho_K.
\end{eqnarray}
In the above equations $J_B$ and $I_{3B}$ are the spin and the
isospin projection and $k_B$ is the Fermi momentum of the baryon
species $B$. The factors in Eqs. (\ref{sigma}) and (\ref{sigmastar}),
$C_B(\sigma)$, $C_B(\sigma^*)$, $C_K(\sigma)$ and $C_K(\sigma^*)$
are, respectively, given by
\begin{eqnarray}
g_{\sigma B} C_B(\sigma) = - \frac{\partial m_B^*}{\partial \sigma},
\label{bary1} \\
g_{\sigma^* B} C_B(\sigma^*) = - \frac{\partial m_B^*}{\partial \sigma^*},
\label{bary2} \\
g_{\sigma K} C_K(\sigma) = - \frac{\partial m_K^*}{\partial \sigma},
\label{ka1} \\
g_{\sigma^* K}C_K(\sigma^*) = - \frac{\partial m_K^*}{\partial \sigma^*}.
\label{ka2}
\end{eqnarray}
Detailed expressions for Eqs. (\ref{bary1}) and (\ref{bary2}) are
given in Refs. \cite{jin,greiner}.
$g_{\sigma K} C_K(\sigma)$ and $g_{\sigma^* K} C_K(\sigma^*)$ for a point-like kaon are nothing but
$g_{\sigma K}$ and $g_{\sigma^* K}$, respectively,
but for a kaon as a MIT bag they need to be calculated self-consistently.
The detailed expressions can be written similarly as in \cite{jin,greiner}.

Let us now apply this model
for hyperonic matter with kaon bags to a neutron star matter.
Neutron star matter is characterized by three conditions;
baryon number conservation,
charge neutrality and chemical equilibrium.
Baryons can be produced when the chemical equilibrium conditions
are satisfied
\begin{eqnarray}
\mu_n = \mu_\Lambda &=& \mu_{\Sigma^0} = \mu_{\Xi^0} , \nonumber \\
\mu_n + \mu_e &=& \mu_{\Sigma^-} = \mu_{\Xi^-} , \nonumber \\
\mu_n - \mu_e &=& \mu_p = \mu_{\Sigma^+},
\end{eqnarray}
where the chemical potential of a baryon is given by
\begin{eqnarray}
\mu_B = \sqrt{k_B^2 + {m_B^*}^2(\sigma,\sigma^*)} + g_{\omega B}\omega_0
+ g_{\phi B} \phi_0 + g_{\rho B} I_{3B}\rho_{03}
\end{eqnarray}
and that of a lepton is simply written as
\begin{eqnarray}
\mu_l = \sqrt{k^2_l + m^2_l}.
\end{eqnarray}
Also, the density of a muon is determined by
$\mu_e = \mu_\mu$.
When kaon condensation takes place, electrons are replaced by $K^-$
so that $n \rightarrow p + K^-$. Therefore, kaons are produced when the condition
$\mu_n - \mu_p = \mu_K$ is met, $\mu_K$ being the chemical potential of a kaon
equal to the kaon energy
in Eq. (\ref{kenergy1}) for S-wave condensation.
Also, the charge neutrality gives us the condition
\begin{eqnarray}
\sum_B q_B \rho_B - \rho_K - \rho_e - \rho_\mu = 0,
\end{eqnarray}
where $q_B$ is the charge and $\rho_B$ is the number
density of baryon species $B$.

The energy density of the matter gets contributions from all the particles,
\begin{eqnarray}
\varepsilon &=& \frac 12 m_\sigma^2 \sigma^2 + \frac 12 m_{\sigma^*}^2 {\sigma^*}^2
+\frac 12 m_\omega^2 \omega_0^2 + \frac 12 m_\phi^2 \phi_0^2 +
\frac 12 m_\rho^2 \rho_{03}^2  \nonumber \\
&&+\sum_B \frac{2J_B+1}{2 \pi^2} \int_0^{k_B} [k^2 + {m_B^*}^2]^{1/2}k^2 dk
+\sum_l \frac{1}{\pi^2} \int_0^{k_l} [k^2 + m_l^2]^{1/2} k^2 dk \nonumber \\
&&+ m_K^* \rho_K,
\label{eq:enden}
\end{eqnarray}
but because there is no contribution to pressure from a kaon in its s-wave
the pressure is given by
\begin{eqnarray}
P &=& -\frac 12 m_\sigma^2 \sigma^2 - \frac 12 m_{\sigma^*}^2 {\sigma^*}^2
+\frac 12 m_\omega^2 \omega_0^2 + \frac 12 m_\phi^2 \phi_0^2 +
\frac 12 m_\rho^2 \rho_{03}^2 \nonumber \\
&&+ \frac 13 \sum_B \frac{2J_B+1}{2 \pi^2} \int_0^{k_B}
\frac{k^4 dk}{[k^2 + {m_B^*}^2]^{1/2}}
+\frac 13 \sum_l \frac{1}{\pi^2} \int_0^{k_l} \frac{k^4 dk}{[k^2 + m_l^2]^{1/2}}.
\label{eq:pressure}
\end{eqnarray}

\subsection{Parameters}
For the bag constant $B$ and $Z_B$ of baryons, we use the parameters given
in Ref. \cite{cyryu07} which are fitted to reproduce the free masses of baryons.
The values of $B_K$ and $Z_K$ for a kaon are determined to fit the free mass of a kaon
with $R_K = 0.4$ fm, $m_{u(d)}=0$ MeV and $m_s = 150$ MeV.
The values are $B_K^{1/4} = 170.752$ MeV and $Z_K = 1.152$.
The coupling constants between u(d) quarks and $\sigma$, $\omega$
and $\rho$ mesons are determined to reproduce nuclear matter properties
at the saturation density ($\rho_0 = 0.17 $ fm$^{-3}$).
The coupling parameters determined for $g_\sigma^u =1.0$
are shown in Table \ref{tab:coupling} taken from Ref. \cite{cyryu07}.
The couplings between the s-quark and $\sigma^*$ and $\phi$
are given by the SU(6) symmetry.
In this work, we assume that the s-quark does not
interact with $\sigma$, $\omega$ and $\rho$ mesons.
\begin{table}
\begin{center}
\begin{tabular}{|c|c|c|c|c|c|c|} \hline
~~$g_\sigma^q$~~  & ~~$g_\omega^q$~~   & ~~${g'}_\sigma^B$~~  &
~~$g_\rho^q$~~    & $m_N^* / m_N$  & $K$ (MeV)  & $a_{\rm sym}$(MeV) \\ \hline
         1.0      & 2.71           & 2.27       &
         7.89     & 0.78           & 285.5      & 32.5      \\ \hline
\end{tabular}
\end{center}
\caption{The coupling constants for $(u,\, d)$-quarks and
$(\sigma,\, \omega,\, \rho)$-mesons in the MQMC
model to reproduce the binding energy $B/A=16$ MeV at the saturation
density $\rho_0= 0.17 {\rm fm}^{-3}$ and symmetry energy $a_{\rm sym}=32.5$ MeV.
$m^*_N/m_N$ and $K$ are the ratio of the effective mass to the free mass of
the nucleon and the compression modulus at the saturation density,
respectively \cite{cyryu07}.}
\label{tab:coupling}
\end{table}

The coupling constants for the interaction between the octet baryons
and mesons can be determined by
using the coupling constants in Table \ref{tab:coupling} and the quark counting rule
\begin{eqnarray}
g_\omega^q &=& \frac 13 g_{\omega N} = \frac 12 g_{\omega \Lambda}
= \frac 12 g_{\omega \Sigma} =g_{\omega \Xi}, \\
g_\rho^q &=& g_{\rho N} = g_{\rho \Sigma} = g_{\rho \Xi},
~~g_{\rho \Lambda}=0, \\
g_\phi^s &=& g_{\phi \Lambda} = g_{\phi \Sigma} = \frac 12 g_{\phi \Xi}
\end{eqnarray}
where
$g_\phi^s = \sqrt 2 g_\omega^{u,d}$ from the SU(6) symmetry.

For coupling constants between the kaon and mesons,
$g_{\omega K}$ and $g_{\rho K}$ can be
obtained by the quark counting rule ($g_{\omega K} = g_\omega^q$ and
$g_{\rho K}=g_\rho^q$), $g_{\sigma^* K}$ can be fixed from $f_0$(980) decay,
and  $g_{\phi K}$ from the SU(6) relation $\sqrt 2 g_{\phi K} = g_{\pi \pi \rho}$.
Then we get $g_{\sigma^* K}=2.65$ and $g_{\phi K}=4.27$.
%We perform the calculation for different values of kaon optical potential
%by treating $g_\sigma'^{B_K}$ as a free parameter.
The value of $g_\sigma'^{B_K}$ is associated with the depths of the
antikaon potential in medium through the relation
\begin{eqnarray}
U_K = -(m_K - m_K^*)-g_{\omega K} \omega_0,
\label{eq:bagkaon_opt}
\end{eqnarray}
and Eqs. (\ref{k-efmas}) and (\ref{kmass}).
With $U_K = -120$, $-140$ and $-160$ MeV and $g_\sigma^q = 1$,
we obtain $g_\sigma'^{B_K}$ as 3.114, 4.317 and 5.615, respectively.

\section{Results}
\begin{figure}
\centering
\includegraphics[width=7.5cm]{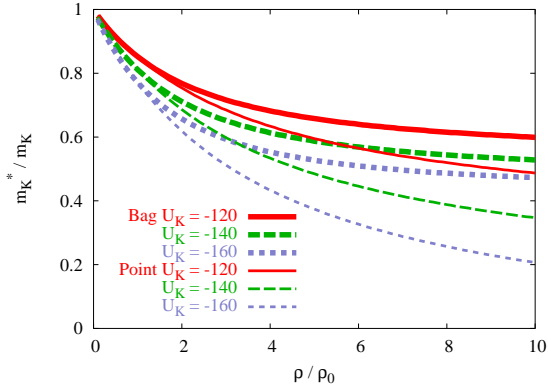}
\caption{The effective mass of a kaon in symmetric nuclear matter is plotted
for both MIT bag kaons (thick curves) and point-particle kaons (thin curves).}
\label{fig-kmas}
\end{figure}

The effective mass of a kaon bag in the MQMC model
in a symmetric nuclear matter is calculated with Eq. (\ref{k-efmas})
and compared with that of a point-like kaon in Fig. \ref{fig-kmas}.
For a point-like kaon, the effective mass is simply given by
$m_K^* = m_K - g_{\sigma K}\sigma$.
For a point-like kaon, $g_{\sigma K}$ is fixed to reproduce the given value
of the kaon optical potential
$U_K = - g_{\sigma K}\sigma - g_{\omega K}\omega_0 $.
Up to $\rho \sim 2 \rho_0$, the effective mass decreases similarly
for both point particle and MIT bag kaons.
For $\rho > 2 \rho_0$, the behaviors are contrasting.
The reduction of the mass of a kaon bag saturates more or less
at high densities, while the mass of a point-like kaon keeps decreasing.

%%%%%%%%%%%%%%%%%%%%%%%%%%%%%%%%%%%%%%%%%%%%%%%%%%%%%%%%%%%%%%%%%%%
\begin{figure}
\centering
\includegraphics[width=5.0cm]{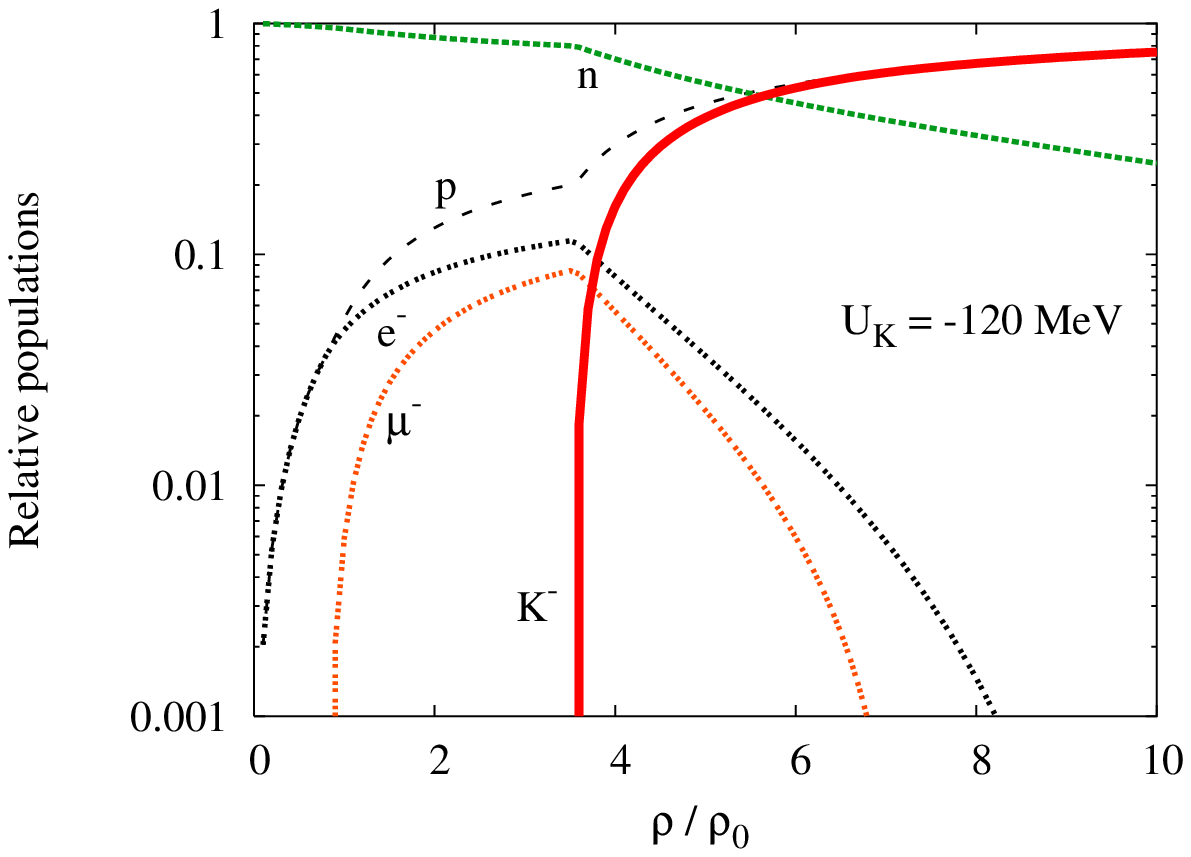}
\includegraphics[width=5.0cm]{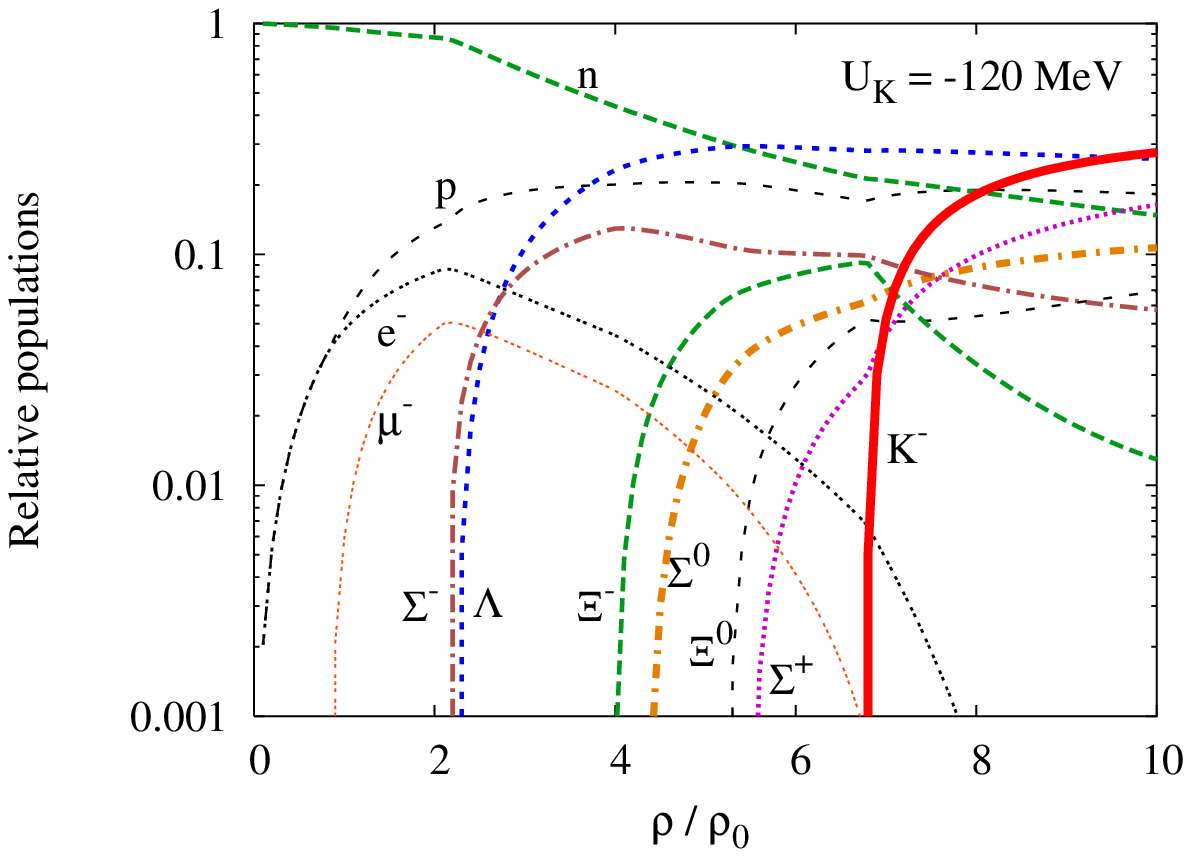}
\includegraphics[width=5.0cm]{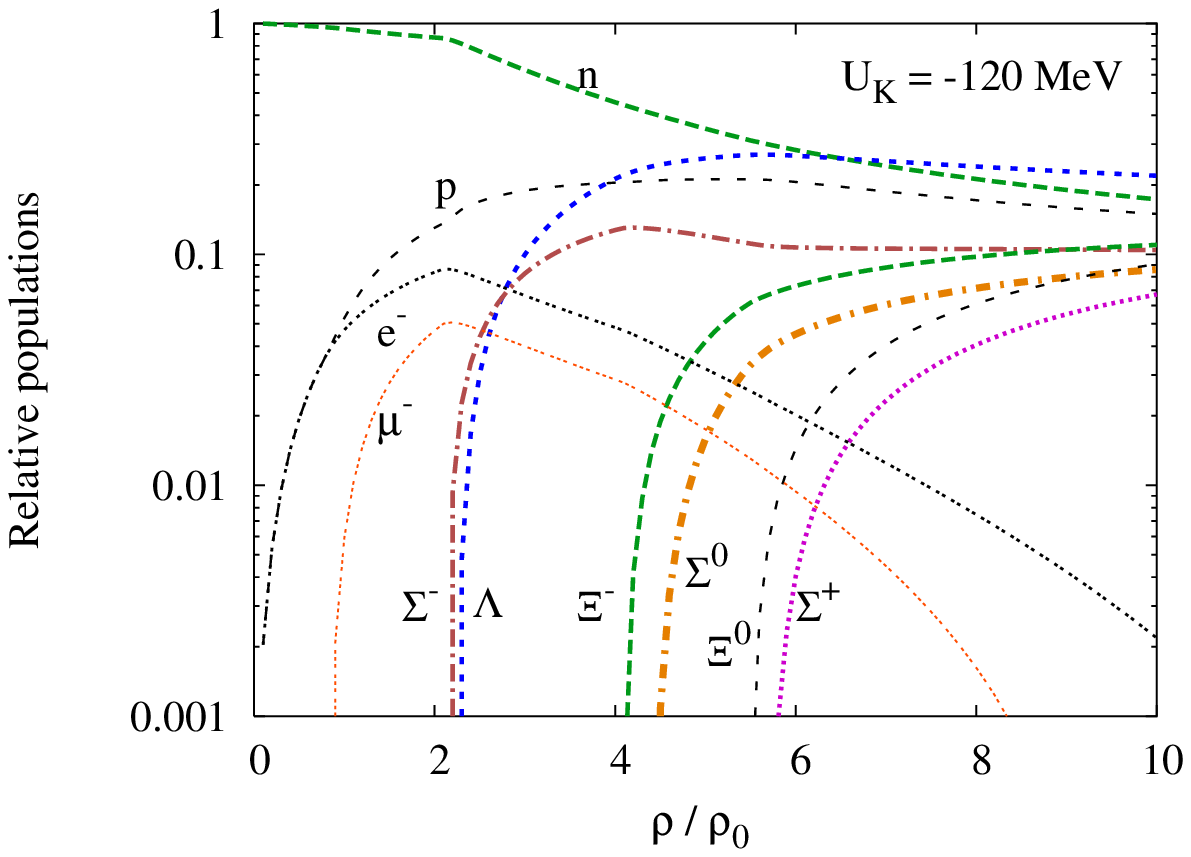} \\
\includegraphics[width=5.0cm]{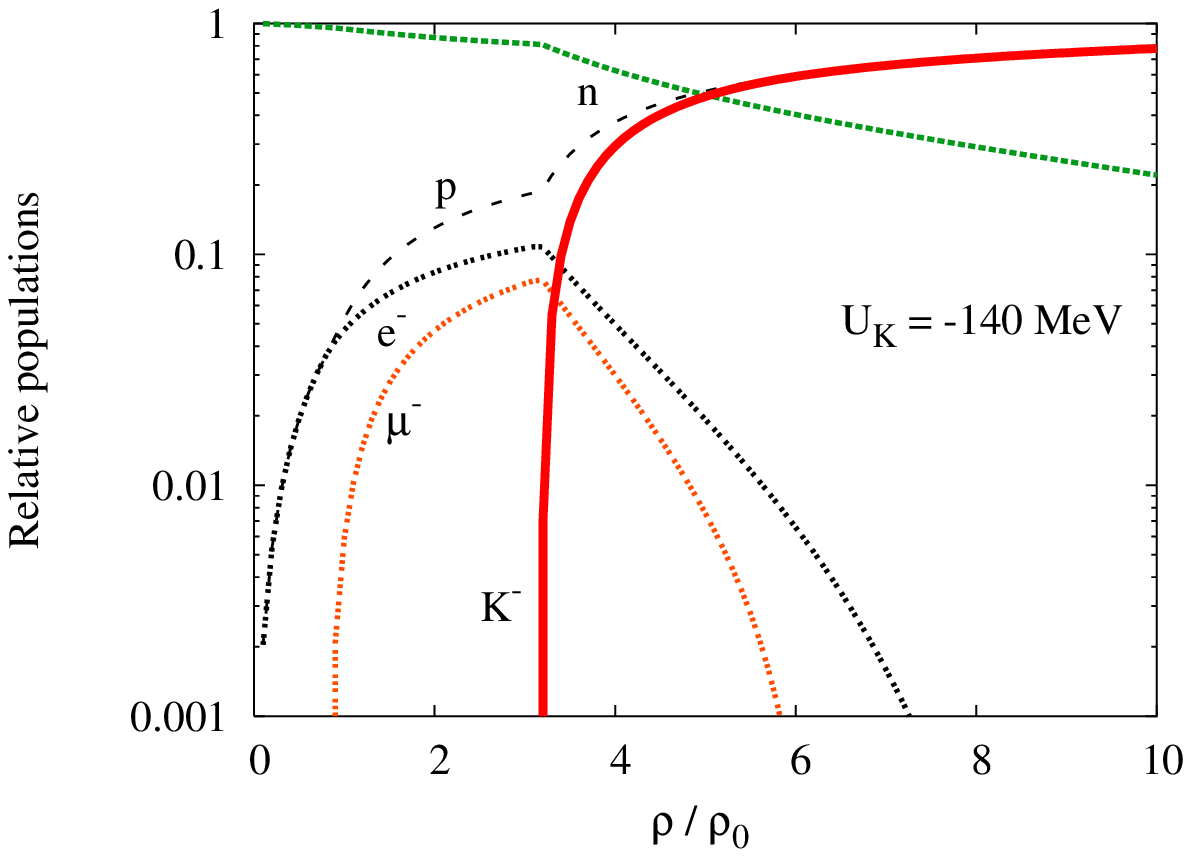}
\includegraphics[width=5.0cm]{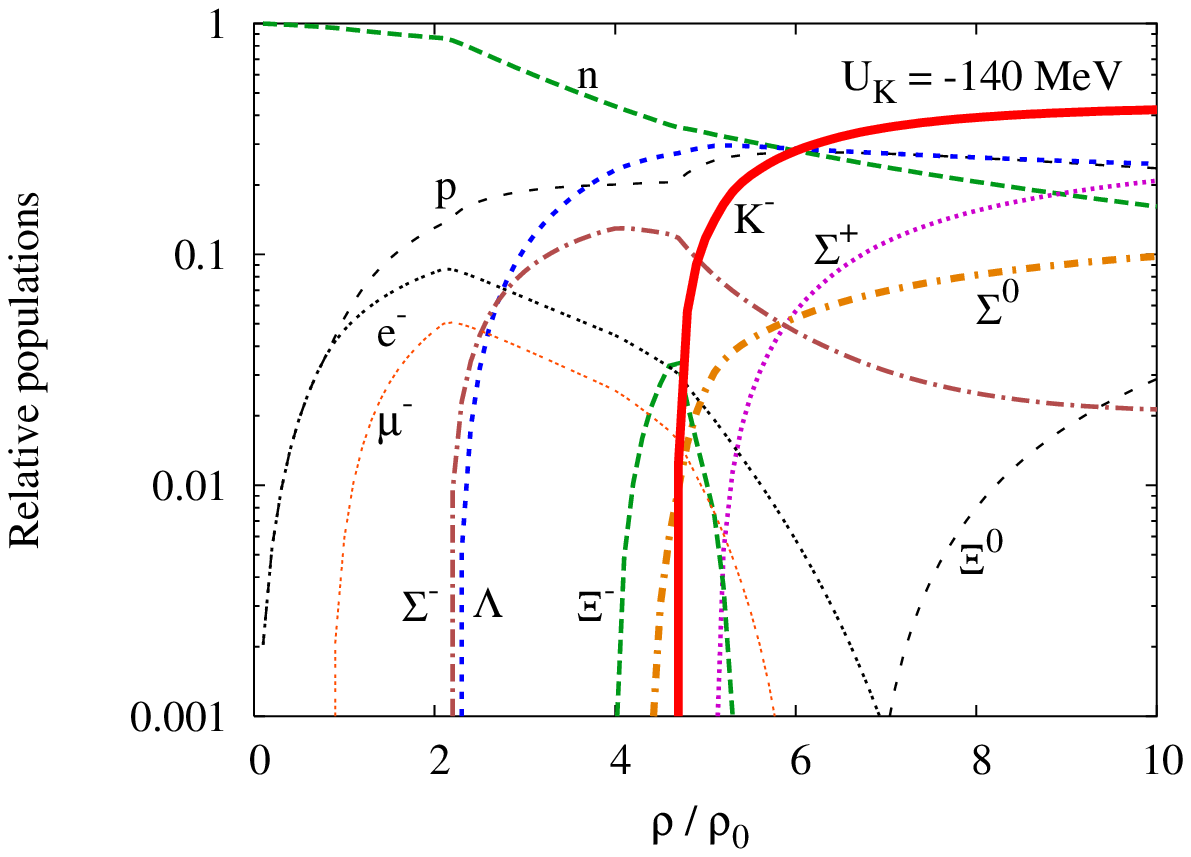}
\includegraphics[width=5.0cm]{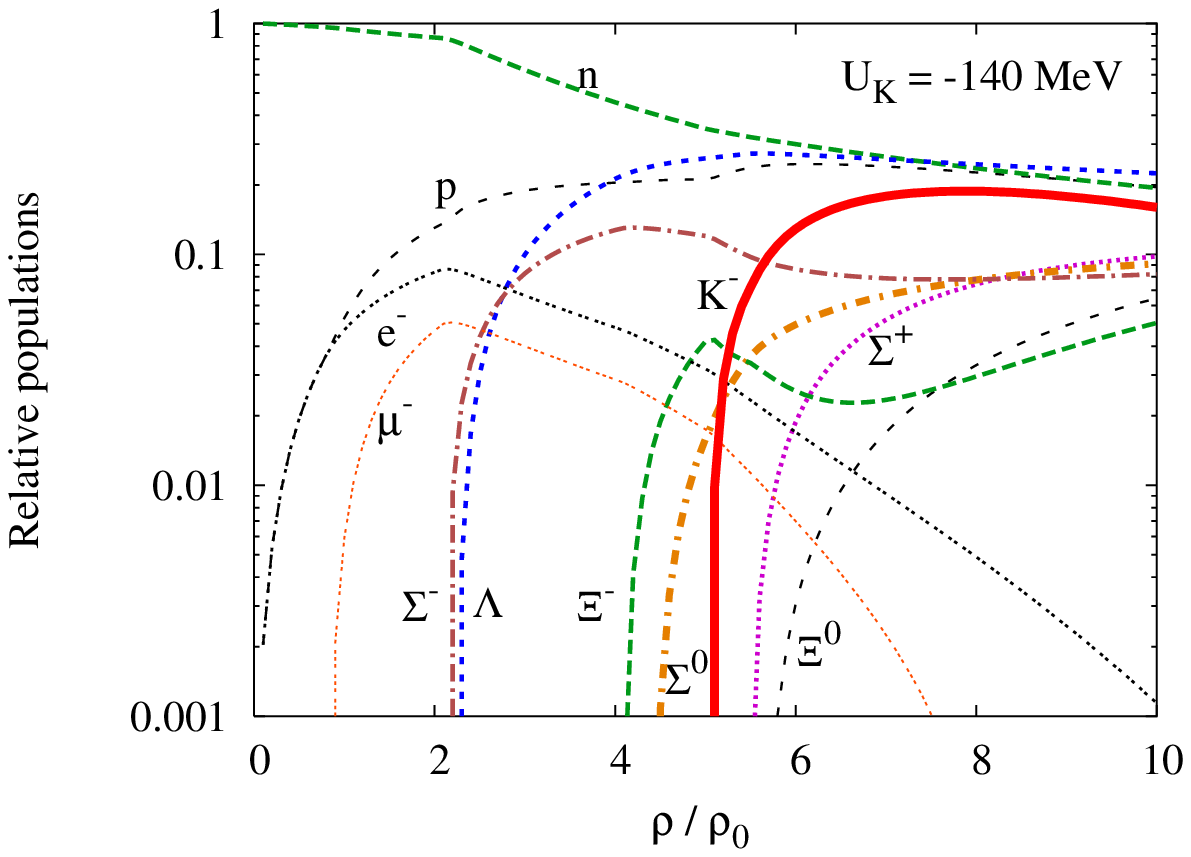}  \\
\includegraphics[width=5.0cm]{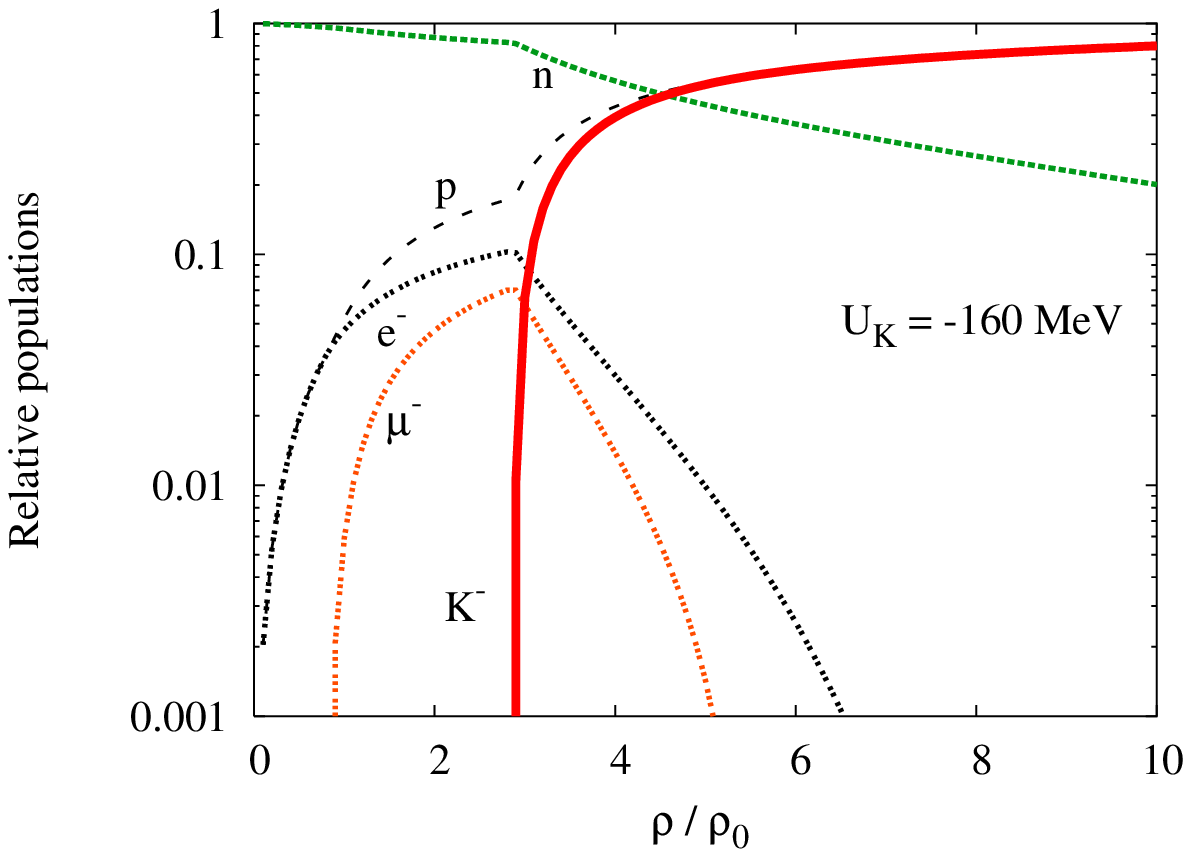}
\includegraphics[width=5.0cm]{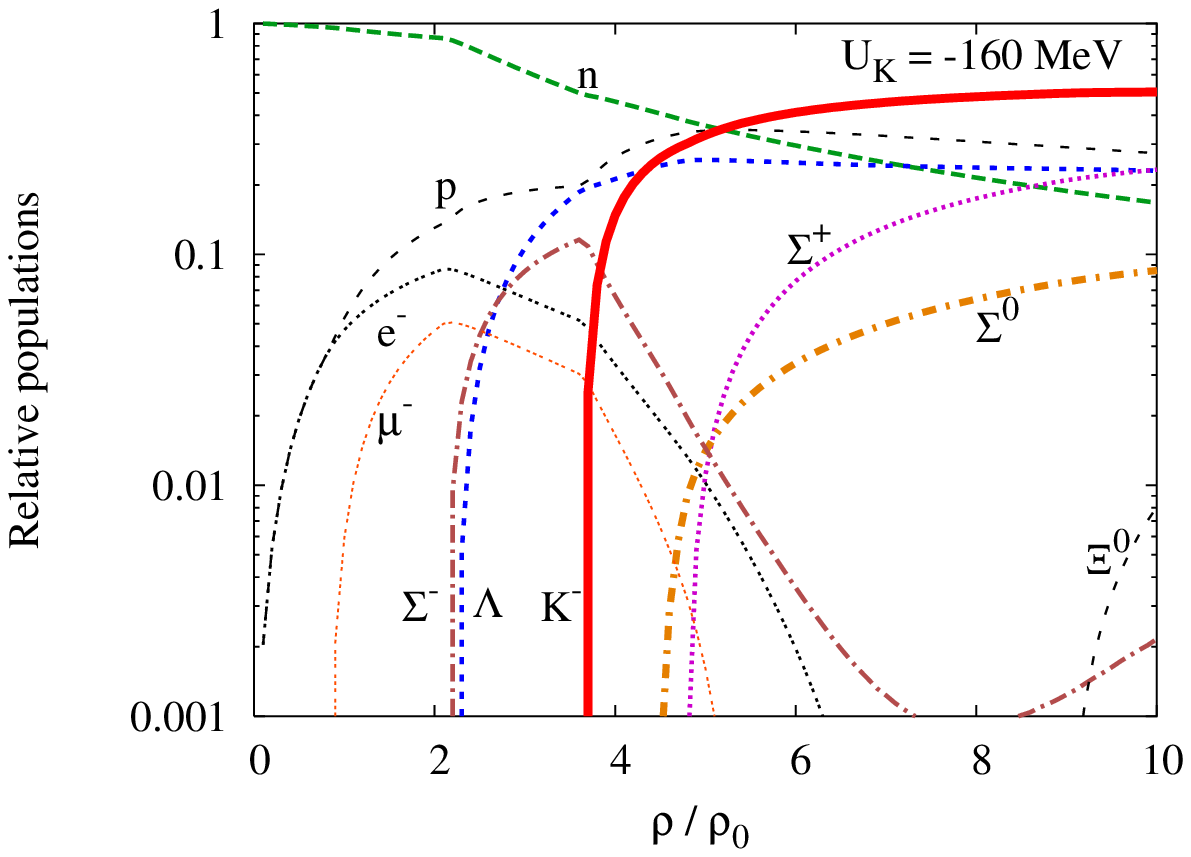}
\includegraphics[width=5.0cm]{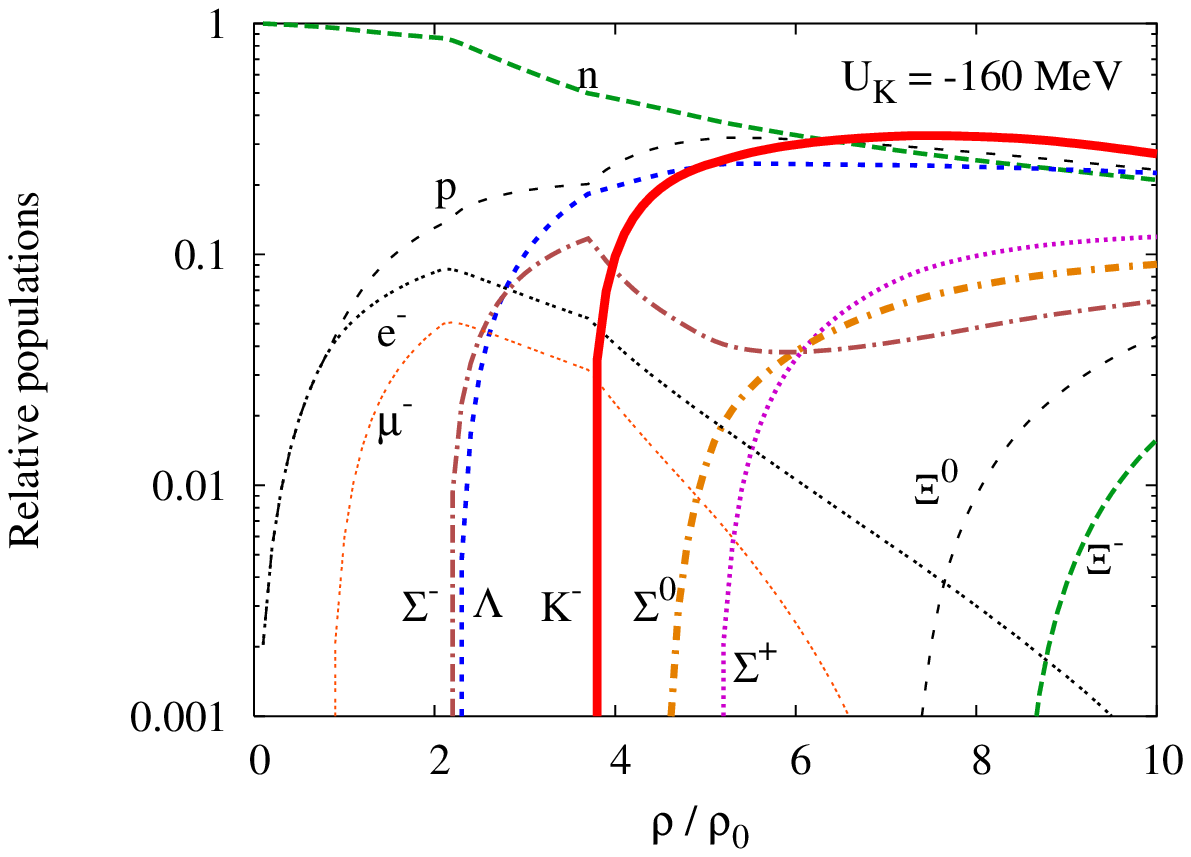}
\caption{The relative populations of particles in nuclear matter (left column),
hyperonic matter without s-quark interactions (middle column),
and hyperonic matter with s-quark interactions (right column)
for the kaon potential $U_K= -120, \,-140, \,-160$ MeV, respectively.}
\label{fig-popul}
\end{figure}
%%%%%%%%%%%%%%%%%%%%%%%%%%%%%%%%%%%%%%%%%%%%%%%%%%%%%%%%%%%%%%%%%%%
\begin{figure}
\centering
\includegraphics[width=5.0cm]{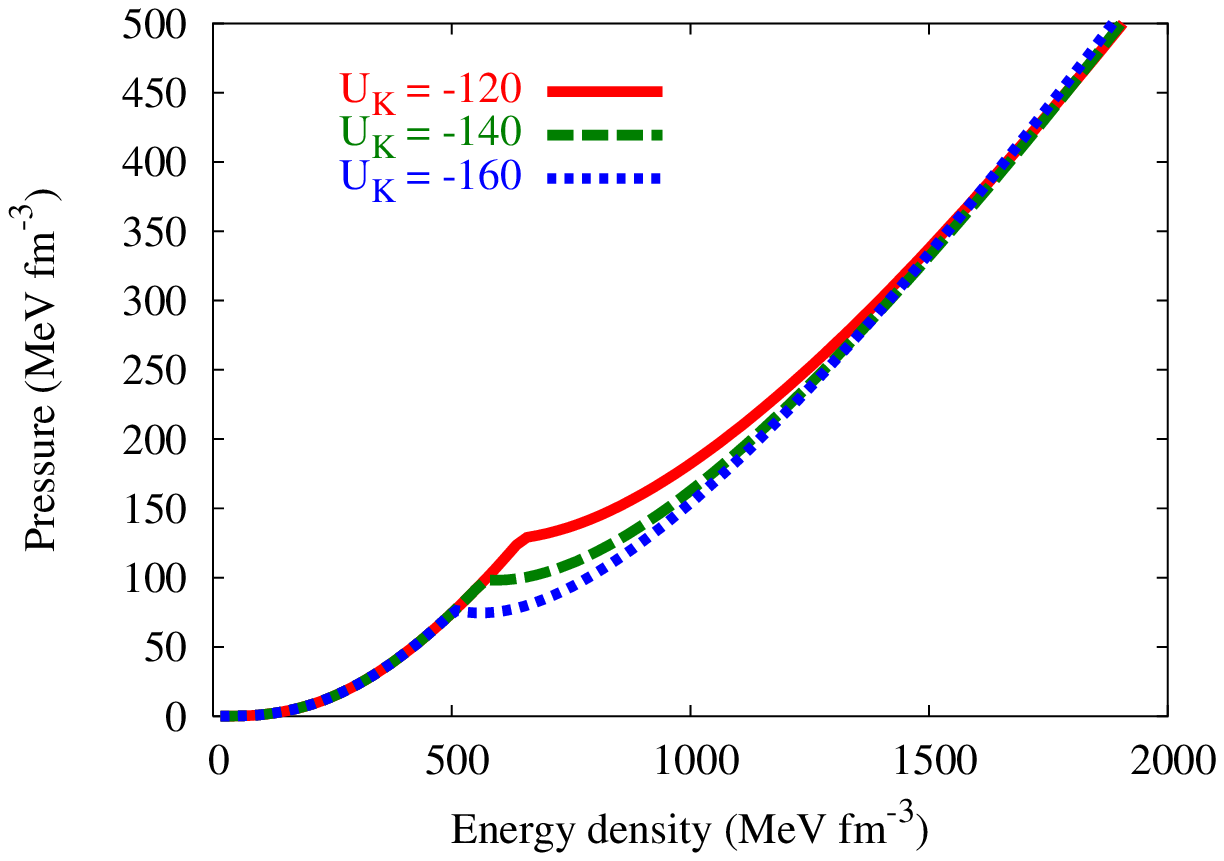}
\includegraphics[width=5.0cm]{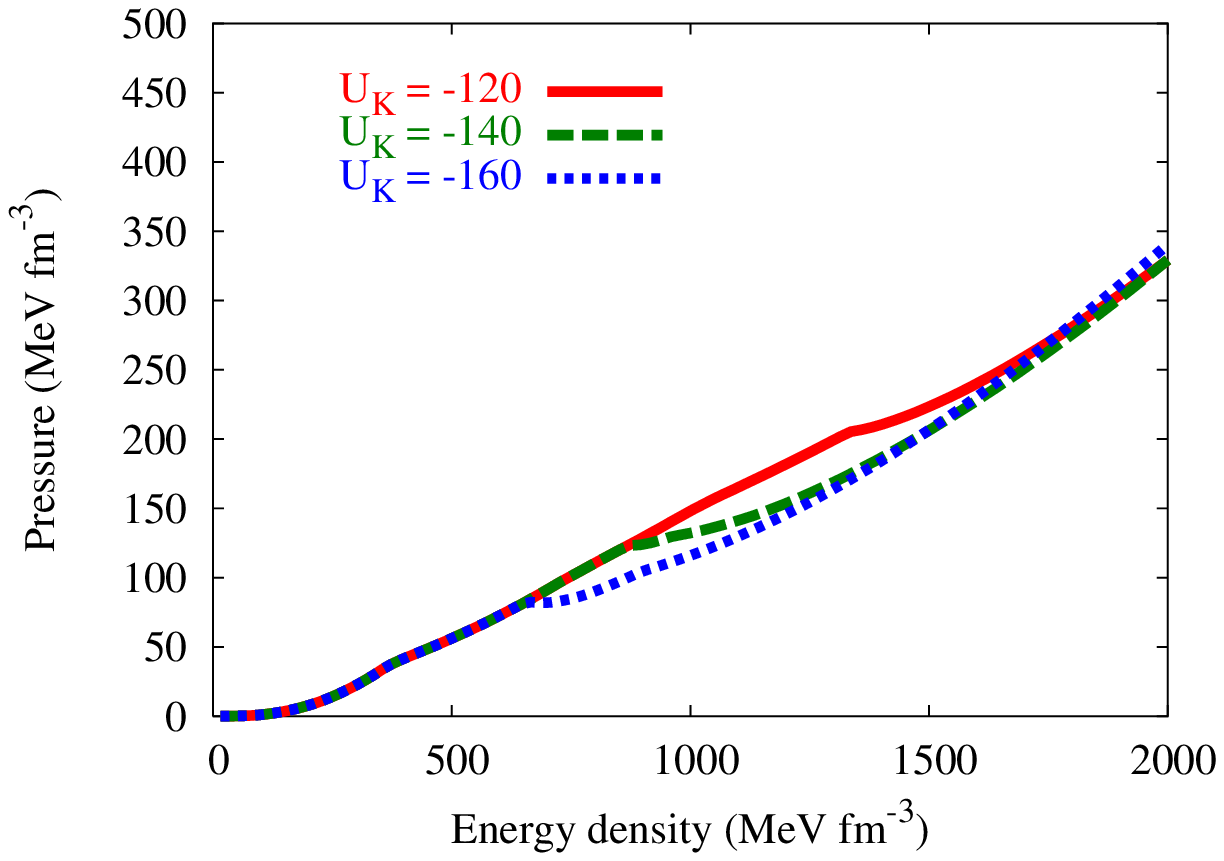}
\includegraphics[width=5.0cm]{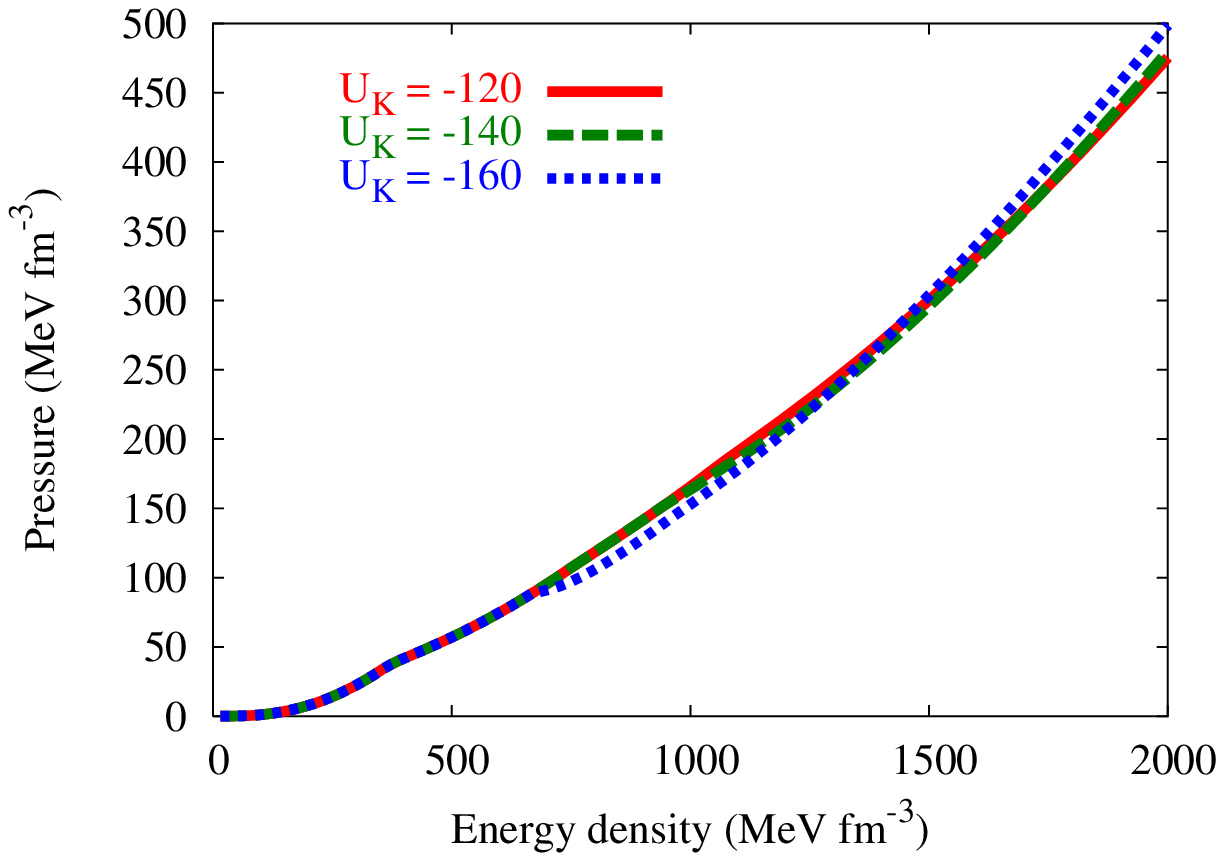}
\caption{The equation of state for nuclear matter with only n, p and K (left),
hyperonic matter (n, p, Y, K) without s-quark interactions (middle),
and hyperonic matter with s-quark interactions (right), respectively.}
\label{fig-eos}
\end{figure}
%%%%%%%%%%%%%%%%%%%%%%%%%%%%%%%%%%%%%%%%%%%%%%%%%%%%%%%%%%%%%%%%%%%

In Figs.~\ref{fig-popul} and \ref{fig-eos},
fractional population of particles and the EoS of neutron star matter
are shown for the matter with nucleons, leptons and kaons ($npK^-$; left column),
for the matter with octet baryons, leptons and kaons without s-quark interaction
(middle column),
and for the matter with octet baryons, leptons and kaons with s-quark interaction
(right column).
Each row in Fig.~\ref{fig-popul} corresponds to $U_K =$ $-120$, $-140$, and $-160$ MeV,
respectively, from the top.

In the case of nuclear matter with $npK^-$,
the onset densities of kaon condensation
are quite low as 3.6, 3.2 and 2.9 times $\rho_0$ for
$U_K$ = $-120$, $-140$ and $-160$ MeV, respectively.
Even the largest onset density $3.6 \rho_0$ is similar to the
mean density of a typical neutron star, mass $\sim 2.0 M_\odot$
and radius $R \sim 10$ km. Thus, if pure nucleonic phase persists
to a certain depth of the neutron star, kaon condensation
is likely to be the state of matter inside the neutron star.
At high densities, population of $K^-$ is dominant, making
the neutron star a kaonic matter.
%Although both $\sigma$ and $\omega$ mesons interact with
%an antikaon attractively,
%$\rho$-meson exchange in asymmetric matter exerts a
%repulsive force to the antikaon, which can prohibits the
%kaon condensation in the neutron star matter.

The EoS for $npK^-$ matter is shown in the left panel of Fig. \ref{fig-eos}.
It is worthwhile to compare our result with that obtained
with point-like kaon in the QHD model \cite{glen-prc}.
In Ref. \cite{glen-prc} the phase transition is treated
as first-order and a very soft EoS is obtained.
On the other hand, because the softening of EoS due to
the kaon condensation is not so substantial in our model,
the phase transition from normal nuclear phase to kaonic phase
depends on how a kaon is treated in medium as well as kaon optical potetial.
However, once the kaon condensation sets in,
the particle composition becomes very similar at high densities
regardless of the $U_K$ value, and it makes the EoS also similar
at high densities.

With hyperon degrees of freedom but without s-quark interactions
($npYK^-$ matter), we obtain the composition of the neutron star matter
shown in the middle column of Fig.~\ref{fig-popul}.
Since antikaon has a negative charge and becomes very dominant once
it is created, all the negatively charged hyperons disappear
soon after the kaon condensation.
Condensation onset densities are 6.8, 4.5 and 3.8 times
$\rho_0$ for $U_K = -120,\, -140$ and $-160$ MeV, respectively.
Dependence of kaon condensation on the value of $U_K$ is stronger
than the $npK^-$ matter case.
The reason for this strong dependence may be partially
attributed to the existence of the hyperons before the kaon
condensation.
As density increases, Fermi momentum of the neutrons reaches to
a value that satisfies the chemical equilibrium for light hyperons
such as $\Lambda$ and $\Sigma^-$.
The creation of the $\Lambda$ and $\Sigma^-$ hyperons reduces
the rate of increase of the neutron Fermi momentum as a function of density,
but the proton Fermi momentum is hardly affected by the hyperons.
Then since $\mu_n - \mu_p$ becomes smaller than that without the
hyperons, kaon creation condition $\mu_K = \mu_n - \mu_p$ becomes
more sensitive to the behavior of the kaon chemical potential.
%Less number of the kaon for $npYK^-$ than that of $npK^-$ can
%be understood in a similar way.}
%

The EoS for $npYK^-$ matter, shown in the middle panel of Fig.~\ref{fig-eos}
is much softer than that for $npK^-$ matter because of the creation of the hyperon.
The softening of the EoS due to kaons on top of the creation of hyperons
is less significant. For instance, the derivative of the EoS $dP/d\varepsilon$
for $npK^-$ matter with $U_K = -160$ MeV abruptly becomes close to zero at the
density where the kaon condensation begins, and the gap between the two
EoS's for $U_K=-120$ MeV and $-160$ MeV is substantial.
On the other hand, the derivative $dP/d\varepsilon$ for $npYK^-$ matter undergoes
only a slight decrease at the density of kaon creation, and so it remains
similar to the value without the kaon condensation.
Kaon condensation plays a significant role in the EoS
if hyperons are not included, but with the inclusion of hyperons
the kaon condensation becomes less significant.

Now, if we switch on the interaction
of s-quarks with $\sigma^*$ and $\phi$ mesons ($npYK^-\phi$)
on top of including the hyperon degrees of freedom,
we get the particle composition in the right column of Fig. \ref{fig-popul}.
When $U_K=-120$ MeV, kaon condensation does not take place at all up to
ten times the saturation density.
For $U_K=-140$ and $-160$ MeV, kaon condensation sets in at
5.1 and 3.8 times $\rho_0$, respectively.
The onset density of the kaon condensation is more sensitive to
the value of $U_K$ than in the $npYK^-$ case.
In addition, the population of the kaon is significantly reduced from
that in the case of $npYK^-$ matter for all the $U_K$ values.
The population of s-quarks increases at higher densities due to
large production of hyperons and kaons.
Thus the values of $\sigma^*$- and $\phi$- fields increase because
strangeness is the source of these meson fields, but at high densities,
$\phi$ meson fields increase more rapidly than $\sigma^*$ \cite{cyryu07}.
Therefore, the repulsion due to $\phi$ mesons inside the hyperon and kaon bags
increase as the population of hyperons and kaons increase.
This repulsion suppresses the kaon condensation.
An interesting thing is that
our results show the repulsion affects the onset of kaon condensation very much,
but the effect is small for the hyperons.
This can be understood by considering the attraction and repulsion
coming from $\sigma$ and $\omega$ fields, respectively.
The net mean field potential of a baryon is a result of the cancelation between
a huge attraction due to $\sigma$ mesons and a huge repulsion due to $\omega$ mesons.
With the s-quark interaction, the net result after the cancelation
between $\sigma^*$ and $\phi$ does not affect much the population of baryons.
For instance, the onset density of $\Xi^-$ with s-quark interaction
changes only slightly from that without the s-quark interaction.
However, in case of the antikaon, both $\sigma$ and $\omega$ mesons
cause only the attraction between baryons and antikaon.
As a result, the effect of repulsion due to s-quark interaction
becomes significant, and in particular, in case of relatively
small potential ($U_K = -120$ MeV), it strongly suppresses kaon condensation.
On the contrary, in case of $U_K = -160$ MeV, the effect of s-quarks
is relatively small to kaon condensation because of the reason mentioned above.

We compare the present result with that of a point-like
kaon in the framework of MQMC \cite{cyryu07}, where
nuclear saturation properties are exactly the same as those in
this work.
If there is any difference, it has to come from how one treats the kaon;
a bag or a point particle.
First, the onset density of the kaon condensation differs
significantly. For the point-like kaon, we have the onset density
5.9, 3.8 and 3.0 times $\rho_0$ for $U_K = -120$, $-140$, and $-160$
MeV, respectively.
Second, the population of the kaon is very different for
the point-like and the bag kaon.
For the point-like kaon, the population of the kaon is even more
dominating than that of the bag kaon without s-quark interactions.
Since the background baryonic matter of the kaon
condensation is built on the same ground with the same saturation
properties, the same models and parameters for baryons and their interactions,
the dramatic difference in the kaon condensation densities
and the kaon population indicates substantial dependency on the
way of how to treat the kaon; a point particle or a bag.
On the other hand, we have observed from the result of the $npYK^-$ case
that if hyperons are included, a softening of the EoS is predominantly
driven by hyperons, and the role of kaons is less significant.
The EoS shown in the right panel of Fig.~\ref{fig-eos} is consistent
with this observation.
The EoS with strange-meson exchange is stiffer than that without
it, which may be ascribed to the extra repulsion due to $\phi$ meson
mean field. The existence of the hyperons, on the other hand, reduces
the pressure substantially, and consequently the EoS for
the $npYK^- \phi$ matter lies in between $npK^-$ and $npYK^-$ cases.

%%%%%%%%%%%%%%%%%%%%%%%%%%%%%%%%%%%%%%%%%%%%%%%%%%%%%%%%%%%%%%%%%%%%%%%
\begin{figure}
\centering
\includegraphics[width=7.5cm]{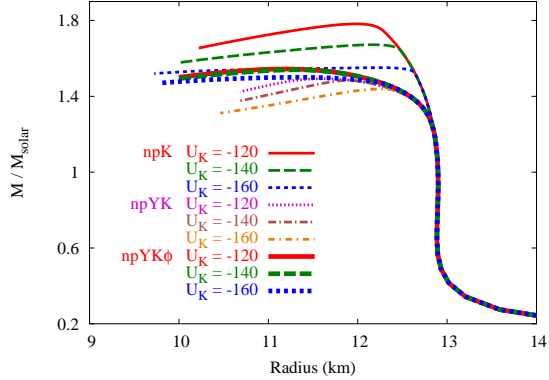}
\caption{The relation of mass and radius of neutron star for $npK^-$,
$npYK^-$ and $npYK^- \phi$.}
\label{fig-mr}
\end{figure}
%%%%%%%%%%%%%%%%%%%%%%%%%%%%%%%%%%%%%%%%%%%%%%%%%%%%%%%%%%%%%%%%%%%%%%%%%%%

Finally, we calculate the mass-radius relation of neutron stars
by solving Tolman-Oppenheimer-Volkoff equation.
The results are plotted in Fig.~\ref{fig-mr}.
The maximum mass of the star for the $npK^-$ case is
about 1.78, 1.67 and 1.55 times the solar mass $M_\odot$
for $U_K = -120$, $-140$ and $-160$ MeV, respectively,
and the corresponding radius is about $12 \sim 13$ km.
Compared with the results in Ref. \cite{glen-prc}, our results yield
the stars with larger mass and radius. In particular, in
Ref. \cite{glen-prc} the radius of the kaonic star is
obtained to be very small (about 8 km) when kaon condensation takes place
with $U_K = - 140$ MeV. Our results show that the mass and
the radius of the star depend much on the model, especially,
on how to treat a kaon in medium.
The maximum masses of the star for the $npYK^-$ matter
are about 1.49, 1.48 and 1.44 times $M_\odot$
for $U_K = -120$, $-140$, $-160$ MeV, respectively.
This model give us a lighter star, which tells us that the effect
of hyperons is very significant.
For the neutron star with $npYK^-\phi$, the maximum mass becomes
about 1.54, 1.54 and 1.50$M_\odot$ for $U_K = -120$, $-140$ and $-160$ MeV.
The stiffer EoS due to the repulsion by the s-quark interaction gives us
larger values of the mass for the case $npYK^-\phi$ than the case of $npYK^-$.
Depending on the value of $U_K$, s-quark interaction can enhance the
maximum mass by $\leq 7$~\%.

The population profile and the EoS of the neutron star depend not
only on the $U_K$ value but also on
whether a kaon is treated as a point particle or a bag.
With a point-like kaon, the maximum mass is in the range
$(1.45 \sim 1.61)M_\odot$ for $U_K = (-160 \sim -120)$ MeV,
an uncertainty exceeding 10\%.
With a kaon bag, on the other hand, we have the maximum mass
$(1.44 \sim 1.49)M_\odot$ for $npYK^-$ and $(1.50 \sim 1.54)M_\odot$
for $npYK^-\phi$, which gives about 3\% uncertainty.

\section{Summary}

We have considered the existence of $K^-$ in dense nuclear
matter by treating it as a MIT bag within the framework
of the MQMC model.
We have investigated the effective mass of $K^-$
in symmetric nuclear matter and
its effect to the properties of the neutron star.
In addition to the standard $\sigma$-, $\omega$- and $\rho$-
meson exchanges which account for the non-strange interactions,
we have also considered the interaction between strange quarks by
including $\sigma^*$- and $\phi$- meson exchanges.
In order to delineate the relative contributions,
we have made various comparisons with and without hyperons,
with and without strangeness interactions,
and from shallow to deep kaon optical potential.
Most importantly, we have compared various results
obtained from a kaon bag and a point-like kaon.

The results without hyperons ($npK^-$) show that kaon condensation
occurs at the density of around $3 \rho_0$, and the dependence
on the kaon optical potential value $U_K$ is relatively weak.
On the other hand, the neutron star mass depends on $U_K$ value
sensitively with the value in the range $M = (1.55 \sim 1.78) M_\odot$ for
$U_K = (-160 \sim -120)$ MeV.
Comparing our $npK^-$ result with the one obtained from the
QHD model in which the nucleon and the kaon are assumed
as point particles \cite{glen-prc}, the results are clearly
contrasting. In the QHD result, kaon condensation depends
sensitively on the $U_K$ value, the EoS in the mixed phase is
much softer than that in this work, and consequently
the maximum mass of the neutron star is 10\% lighter
than our results.
Qualitatively, however, QHD and MQMC have a common aspect:
the effect of the kaon condensation to the matter composition,
the EoS and the neutron star properties is significant and dominant
when there is no hyperons.

If we include the hyperons, $\Lambda$ and $\Sigma^-$ hyperons are
created at a density of about $2.2 \rho_0$.
The creation of the hyperons brings down the highly accumulated neutron's
Fermi levels, and it causes the role of the kaon to be
minor in many respects.
The onset density of the kaon condensation is more sensitive to
the $U_K$ value than that without hyperon,
and the population of the kaon is much less than that for the $npK^-$ case.
The softening of the EoS due to kaon condensation is non-negligible
even with the hyperon, but it is not so significant as that
for the $npK^-$ matter.
The neutron star mass differs by about 3\% depending on $U_K$ value.
This weak dependence on $U_K$ value is in contrast to that
for a point-like kaon, which is about 10\% \cite{cyryu07}.
With the strangeness interaction, repulsion due to
$\phi$-meson exchange suppresses the role of the kaon further.
As a result, the EoS for the $npYK^-\phi$ matter is almost the same as
that without the kaon condensation for all the $U_K$ values considered.
The maximum mass fluctuates by about 3\% depending on the $U_K$ value,
which may be within the uncertainties due to the model dependence,
input parameters such as nuclear saturation properties,
nucleon-nucleon, nucleon-hyperon and hyperon-hyperon interactions,
and etc.
In the case of $npYK^-$ and $npYK^-\phi$ matter,
compared to the point-like kaon, the role of the bag kaon is
quite suppressed and the difference in the mass of the neutron
star is within the errors due to various uncertainties.

%Investigations of the kaon condensation in the neutron star matter
%for more than a decade lead us to draw the following conclusions:
%1) Kaon condensation is sensitive to the model and the input parameters.
%2) If hyperon is included, the effect of the kaon is
%sub-leading correction to the dominant contribution from the octet baryons.

\section*{Acknowledgments}

The work was supported by the Korea Research Foundation Grant
funded by the Korean Government(MOEHRD) (KRF-2006-214-C00015)
and by the Korea Science and Engineering Foundation grant funded
by the Korean Government (MEST) (No. M20608520001-08B0852-00110).


\begin{thebibliography}{99}
\bibitem{akaishi} Y. Akaishi and T. Yamazaki, Phys. Rev. C {\bf 65}, 044005(2002).
\bibitem{suzuki} T. Suzuki {\it et al}., Phys. Lett. B {\bf 597}, 263 (2004).
\bibitem{sato08} M. Sato {\it et al}., Phys. Lett. B {\bf 659}, 107 (2008).
\bibitem{finuda} M. Agnello {\it et al}., Phys. Rev. Lett. {\bf 94}, 212303 (2005).
\bibitem{oset} E. Oset and H. Toki, Phys. Rev. C {\bf 74}, 015207 (2006).
\bibitem{magas} V. K. Magas, A. Ramos, E. Oset and H. Toki,
Phys. Rev. C {\bf 74}, 025206 (2006).
\bibitem{kaplan} D. B. Kaplan and A. E. Nelson, Phys. Lett. B {\bf 175}, 57 (1986).
\bibitem{brown} G. E. Brown and M. Rho, Phys. Rev. Lett. {\bf 66}, 2720 (1991).
\bibitem{brown2} G. E. Brown, Chang-Hwan Lee, M. Rho and V. Thorsson,
Nucl. Phys. {\bf A 567}, 937 (1994).
\bibitem{muto} T. Muto, Phys. Rev. C {\bf 77}, 015810 (2008).
\bibitem{glendening} N. K. Glendenning, J. Schaffner-Bielich,
Phys. Rev. Lett. {\bf 81}, 4564 (1998).
\bibitem{glen-prc} N. K. Glendenning, J. Schaffner-Bielich,
Phys. Rev. C {\bf 60}, 025803 (1999).
\bibitem{banik} S. Banik and D. Bandyopadhyay, Phys. Rev. C {\bf 64}, 055805 (2001).
\bibitem{menezes} D. P. Menezes, P. K. Panda and C. Providencia,
Phys. Rev. C {\bf 72}, 035802 (2005).
\bibitem{cyryu07} C. Y. Ryu, C. H. Hyun, S. W. Hong and B. T. Kim,
Phys. Rev. C {\bf 75}, 055804 (2007).
\bibitem{guichon} P. A. M. Guichon, Phys. Lett. B {\bf 200}, 235 (1988).
\bibitem{fleck} S. Fleck, W. Bentz, K. Shimizu, and Yazaki,
Nucl. Phys. {\bf A 510}, 731 (1990).
\bibitem{jin} X. Jin and B. K. Jennings, Phys. Rev. C {\bf 54},
1427 (1996).
\bibitem{saito-npa} K. Tsushima, K. Saito, J. Haidenbaur and A. W. Thomas,
Nucl. Phys. A {\bf 630}, 691 (1998).
\bibitem{saito98} K. Tsushima, K. Saito, A. W. Thomas and S. V. Wright,
Phys. Lett. B {\bf 429}, 239 (1998).
\bibitem{cyryu05} C. Y. Ryu, C. H. Hyun, J. Y. Lee and S. W. Hong,
Phys. Rev. C {\bf 72}, 045206 (2005).
\bibitem{greiner} S. Pal, M. Hanauske, I.Zakout, H. Stocker and W. Greiner,
Phys. Rev. C {\bf 60}, 015802 (1999).

\end{thebibliography}
\end{document}